\documentclass[runningheads]{llncs}
\usepackage[english]{babel}
\usepackage[T1]{fontenc}
\usepackage[letterpaper,top=2cm,bottom=2cm,left=3cm,right=3cm,marginparwidth=1.75cm]{geometry}
\usepackage{amsmath}
\usepackage{graphicx}
\usepackage[colorlinks=true, allcolors=blue]{hyperref}
\usepackage{comment}

\begin{document}

\title{Semantic Preprocessing for LLM-based Malware Analysis}

\author{Benjamin Marais \and
Tony Quertier \and
Grégoire Barrué}

\institute{Orange Innovation, Cesson-Sevigne, France\\
\email{benjamin.marais@orange.com}\\
\email{tony.quertier@orange.com}\\
\email{gregoire.barrue@orange.com}}

\maketitle

\begin{abstract}

In a context of malware analysis, numerous approaches rely on Artificial Intelligence to handle a large volume of data. However, these techniques focus on data view (images, sequences) and not on an expert's view. Noticing this issue, we propose a preprocessing that focuses on expert knowledge to improve malware semantic analysis and result interpretability. We propose a new preprocessing method which creates JSON reports for Portable Executable files. These reports gather features from both static and behavioral  analysis, and incorporate packer signature detection, MITRE ATT\&CK and Malware Behavior Catalog (MBC) knowledge. The purpose of this preprocessing is to gather a semantic representation of binary files, understandable by malware analysts, and that can enhance AI models' explainability for malicious files analysis. Using this preprocessing to train a Large Language Model for Malware classification, we achieve a weighted-average F1-score of 0.94 on a complex dataset, representative of business reality.

\keywords{Malware Analysis\and
Data Preprocessing\and
Large Language Model\and
Cybersecurity.}
\end{abstract}

\section*{Introduction}

While Artificial Intelligence is now an usual tool in the domain of Malware detection and Malware classification \cite{Gaber/Ahmed}, the training of the models is still one of the main issues, because of the representation of the data. Some works try to solve this problem by extracting more information from the samples, like EMBER \cite{anderson2018EMBER}, while other works try to change the representation of the files, for instance by transforming them into images \cite{nataraj2011malware}, in order to train a convolutional neural network, or directly by analyzing the byte sequence of the binary file \cite{raff2017malwaredetectioneatingexe}. In the same way as other feature extraction methods \cite{sun2017malware}, EMBER provides lots of features for each sample. However, it can be hard to understand which are the most impactful \cite{oyama2019identifying}. There are other techniques for feature extraction \cite{el2022comparison}, some of them coupling static with behavioral analysis \cite{zhang2019feature}, but they all seem to suffer from the fact that they are not easy to understand, and thus do not allow some result explainability. Note that EMBER recently released a new version, providing more detailed reports in the PE file format \cite{EMBER2024}.   

This lack of relevant representation results in a bad classification of malware categories, and a poor explainability of the results. We need to add to these reports more contextual information. The aim is to get benefits from the summarization capabilities of the Large language Models to extract the most important information from the reports. Indeed, Large Language Models (LLM) offer a new opportunity in lots of domains, and Malware detection and classification is obviously one of them. The ability of these models to learn from semantic reports, coupled to the attention mechanism, is very useful. LLMs are used for instance to try to understand code \cite{wang2023codet5+}, or even to disassemble obfuscated code \cite{rong2024disassembling}.    

In this article, we want to provide a new preprocessing for PE file feature extraction, which uses both static and behavioral analysis in order to create a JSON report of a PE file sample, containing detailed information. We illustrate the relevance of this preprocessing by using it to classify Malware categories thanks to LLM models. This classification task is actually more complex than the usual detection task, and some works try for instance to use image classification \cite{khadilkar2024imagebasedmalwareclassificationusing}, or the features extracted from EMBER \cite{EMBER2024}, \cite{loi2021automatedpipelinedetectingclassifying}. While EMBER provides a standardized set of static PE features for malware classification, it does not include any contextual information derived from known rule-based threat signatures. Our approach complements existing datasets (\cite{harang2020sorel20mlargescalebenchmark}, \cite{joyce2023maldictbenchmarkdatasetsmalware}, ...) by incorporating both packer signature detections and YARA rule matches. These additional features aim to capture higher-level semantic signals and enhance interpretability, which are crucial for threat hunting and advanced malware triage. One advantage of our preprocessing is that the reports can be understood directly by an analyst, since they contain concrete information.
This article is organized as follows. In Section \ref{sec:Methodology} we present our dataset, the format of our files and the preprocessing that we use to create our reports. Then, Section \ref{sec:Experiments} gathers information about the experiments that we run with our reports, by training two Large Language Models for our classification task, and details the results that we obtained. Finally, we conclude about the advantages of using our complete preprocessing to train a LLM and discuss about future works.





\section{Methodology}
\label{sec:Methodology}

In this section, we present the methodology of our paper by describing the different fundamental blocks. To begin with, we detail the PE format and the dataset used in our work. Then, we describe the preprocessing proposed for our experimentation with the features that we selected and how we extract them from the binary files. 

\subsection{File format and dataset}

\subsubsection{Portable Executable Format}

Malicious files remain an important cybersecurity threat, accounting for over 60\% of successful cyberattacks in 2024 \cite{global_ptsecurity,kapersky_2024}. Among them, malware based Portable Exectuable (PE) format represents more than $95\%$ of malicious files in circulation \cite{avtest2025}, and are commonly use by threat actor to hide and diffuse malicious payload and commit exploits \cite{mendiant_pe_malware_2024}.

\begin{figure}[!h]
	\centering
	\includegraphics[width=.7\linewidth]{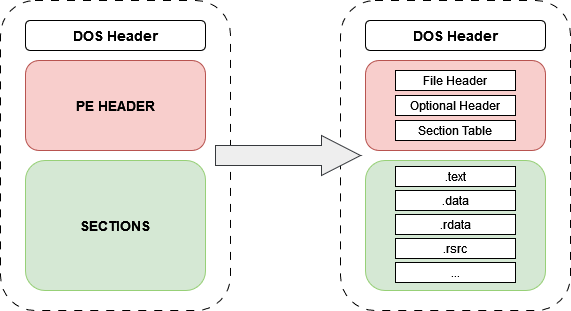}
	\caption{Simplified architecture of a PE file}
	\label{fig:PEFormat}
\end{figure}

The PE format is a file format inherited from the COFF (Common Object File Format) format mainly used by UNIX systems. This format was created in 1993 by Microsoft \cite{microsoft_pe_format_2002} to be used with Windows operating system files. It is mainly used for DLLs (.dll) and executable (.exe) files. As shown in Figure \ref{fig:PEFormat}, a PE file is divided into two parts: the header and the sections. The header describes the file and its contents. It contains, among other things, information such as the file creation date, information on file loading into memory and the number of sections. Each section is described by a specific header that includes its name, size and location in virtual memory. Sections generally contain the executable code (.text), the variables used and their default values (.data) or information on the functions (DLLs) required to execute the file (.idata).

By studying the information contained in a file's header and sections, as well as its behavior (and purpose), it is possible to determine whether a file is potentially malicious or not.

\subsubsection{Dataset}

The files used for our work come from an open source dataset, available online. We selected the BODMAS dataset \cite{yang2021bodmas} for malicious files. A particularity of this dataset is that the types (or categories) of the files are described in the metadata provided by the authors. In this dataset we isolated 8 categories, which were the most represented in this dataset, the others not having enough samples to train a model to classify them. The different categories are: trojan, worm, ransomware, backdoor, information stealer, downloader, dropper, and virus. They differ by their behavior once a computer is infected; for example a ransomware will encrypt the computer's data to force the owner to pay, whereas a trojan can hide its malicious behavior in order to steal data from the infected computer. 

We chose to use a relatively big dataset (around 57,000 samples) compared to some other works \cite{nataraj2011malware,ronen2018microsoftmalwareclassificationchallenge}, in order to ensure a good training of our model. This involves having under-represented malware categories because it is sometimes harder to find some types of malware and easier to find some other types, but in our opinion this allows to have a realistic distribution of the different categories. Besides, if we wanted to have a perfectly balanced dataset, we would have been forced to drastically reduce the size of our dataset, thus deteriorating the quality of the training.  
\begin{table}[!h]
    \centering   
    \caption{BODMAS' malware repartition}
    \label{tab:rep_bodmas}
    \begin{tabular}{|c|c|} \hline
        File Categories & Number of instance \\ \hline\hline
        Trojan & $29,972$   \\ \hline
        Worm & $16,697$   \\ \hline
        Backdoor & $7,331$   \\ \hline
        Downloader & $1,031$   \\ \hline
        Ransomware & $821$   \\ \hline
        Dropper & $715$   \\ \hline
        Infostealer & $448$   \\ \hline
        Virus & $192$   \\ \hline
    \end{tabular}
\end{table}

\subsection{Preprocessing and Features Description}

Based on our malware analysis experience, on expert knowledge, and inspired by work as EMBER \cite{anderson2018EMBER}, we propose the following preprocessing to convert a binary file into a convenient form to use it with machine learning or deep learning models. We extract substantial information from the file, generally used for static malware detection and analysis, and format this information into a report as a JSON file. As EMBER, we mainly focus on extracting static features contained in binary files, because they are the main characteristics used by malware analysts in this field. However, we also extract information based on rules or signatures to enrich our JSON reports. These rules or signatures give behavioral (MITRE ATT\&CK and MBC) or obfuscation (packing) information. In the remainder of this section, we will describe in more details the characteristics extracted from the binary files. For each subsection of the JSON report, we will provide examples based on a variant of the Wannacry ransomware \cite{wannacry}. 

\begin{figure}[!h]
    \centering
    \includegraphics[width=0.7\linewidth]{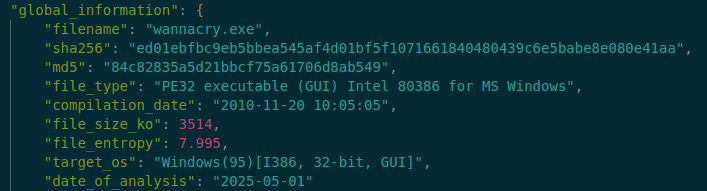}
    \caption{Global file information}
    \label{fig:global_section}
\end{figure}

The first section of a JSON report provides \textbf{global information} on the analysis file as the file name, signature (sha256, md5) the file type, the target OS, the compilation date, the file size and the file entropy. This information describes the file, and for instance a weird compilation date or a hight entropy can be indicators of a suspicious behavior. However, it cannot be used to decide whether a file is a benign file or a malware, and it needs to be completed with all the information explained below. Figure \ref{fig:global_section} provides an example if such a JSON section.
 
The second part of the report gives information related to the \textbf{sections} of the file. In PE files, the section table describes how the program is organized in memory. Each section contains data or code needed during runtime or linking. A common section includes \textit{.text} (executable code), \textit{.data} (initialized data), \textit{.rdata} (read-only data) and \textit{.rsrc} (resources like icons or dialog). Each section has a set of characteristics, with a flag describing the section's attributes, such as whether it is executable, readable, writable or contains code, as described in Table \ref{tab:table-sections}. 

\begin{table}[!h]
    \centering
        \caption{Some section characteristics}
    \label{tab:table-sections}
    \begin{tabular}{|c|c|}\hline
         Flag Name & Description \\\hline\hline
         MEM EXECUTRE  &  Section contains executable code\\\hline
         MEM WRITE  & Section is writable\\\hline
         MEM READ  &  Section is readable.\\\hline
         CNT CODE  &  	Section contains code\\\hline
         CNT INITIALIZED DATA  &  Section contains initialized data\\\hline
    \end{tabular}
\end{table}

\begin{figure}[!h]
    \centering
    \includegraphics[width=0.7\linewidth]{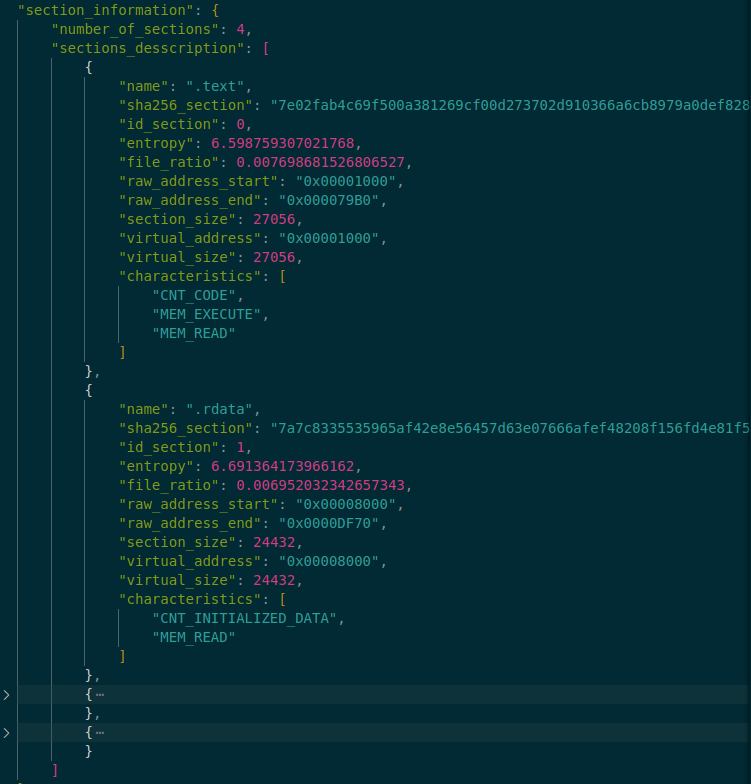}
    \caption{File section information}
    \label{fig:section_section}
\end{figure}

It is common to base malware analysis on section analysis \cite{quertier2024use,XIAO2021102420}. For example, changing section names is a common technique used to evade detection or hinder reverse engineering \cite{anderson2018learningevadestaticpe,quertier2022merlinmalwareevasion}. In the other hand, some section names are a characteristic of evasion tools such as UPX\footnote{https://upx.github.io/} that compress the original code and data into custom sections like \textit{.upx1} or \textit{.upx2}. Additionally, malware may assign unusual characteristics to a section, such as making the \textit{.rsrc} executable, which may indicate a hidden malicious payload in this section. These irregularities, in naming and characteristics, are red flags and often indicate packing, obfuscation or code injection techniques used to bypass antivirus engines or harden reverse engineering. In our JSON report, as described in Figure \ref{fig:section_section}, we keep, among other things, the name, the size, the signature (sha256) and the characteristics of the sections. 

\begin{figure}[!h]
    \centering
    \includegraphics[width=0.7\linewidth]{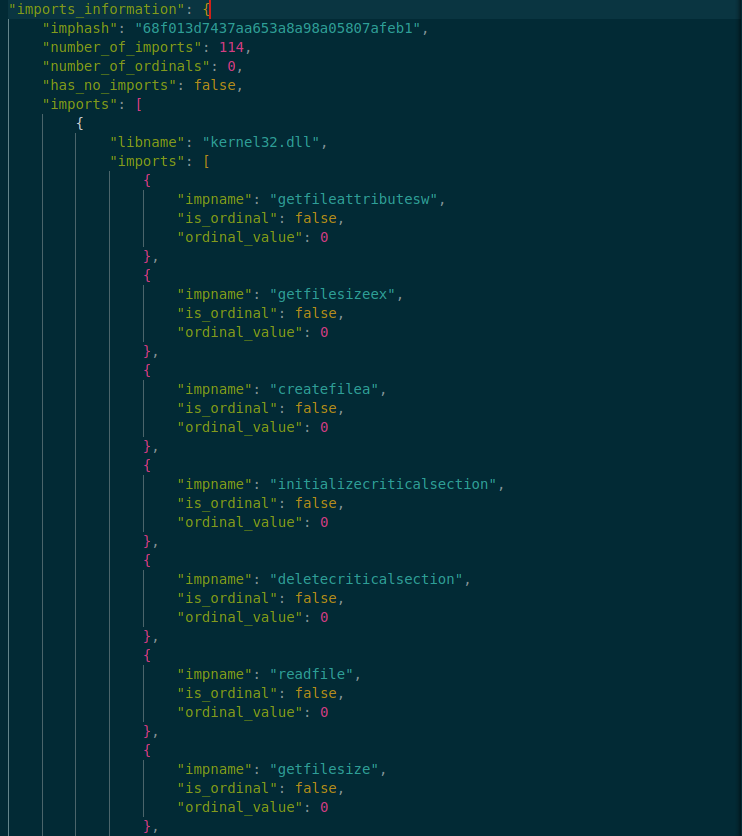}
    \caption{IAT information}
    \label{fig:imp_section}
\end{figure}

The \textbf{Import Address Table (IAT)} is a critical structure within a PE (Portable Executable) file that lists all the external functions a program relies on, typically from Windows DLLs like kernel32.dll, user32.dll, or advapi32.dll. Each entry in the IAT corresponds to a function that will be resolved and loaded by the Windows loader at runtime, pointing to the actual memory address of the function in the relevant DLL. The IAT contains both the names of the imported functions and the DLLs they come from. In our preprocessing, we extract information related to the IAT. As described in Figure \ref{fig:imp_section}, we first give the imphash, a file signature related to the imported functions, and the number of clear and ordinal imported functions. Then we give the list of all imported functions, sorted by library (kernel32.dll for example). If a function is imported by ordinal, we retrieve, if possible, its name.

During static analysis, the IAT provides valuable insight into a binary’s behavior. For instance, imports by ordinal instead of by name can be suspicious \cite{ordinal}, as it is a common tactic used to evade detection or obfuscate a malicious intent. Analysts also look for malicious patterns in imported functions, such as the presence of API calls often used in code injection or process hollowing, for example.  A cluster of the imports contained can strongly suggest the binary may be performing one of the actions described in Table \ref{tab:table-import}. Therefore, scrutinizing the IAT is a key step in identifying potential threats during static malware analysis.

\begin{table}[!h]
    \centering
        \caption{Some risky windows API calls from \cite{zeltser}}
    \label{tab:table-import}
    \begin{tabular}{|c|c|}\hline
    Exploit  & Windows API Calls  \\ \hline\hline
    Code injection & CreateRemoteThread, OpenProcess,  \\ 
                   & VirtualAllocEx, WriteProcessMemory, \\
                   & EnumProcesses\\ \hline
    Dynamic DLL loading & LoadLibrary, GetProcAddres \\ \hline
    Memory scrapping & CreateToolhelp32Snapshot, OpenProcess \\
                     & RadProcessMemory, EnumProcess \\ \hline
    Unpacking/self-injection & ViruatlAlloc, VirtualProtect \\ \hline
    Execute a program & WinExec, ShellExecute, \\
                      & CreateProcess \\ \hline
    Query artifact & CreateMutex, CreateFile \\
                   & FindWindow, GetModuleHandle, \\
                   & RegOpenKeyE \\\hline            
    \end{tabular}
\end{table}

\begin{figure}[!h]
    \centering
    \includegraphics[width=0.7\linewidth]{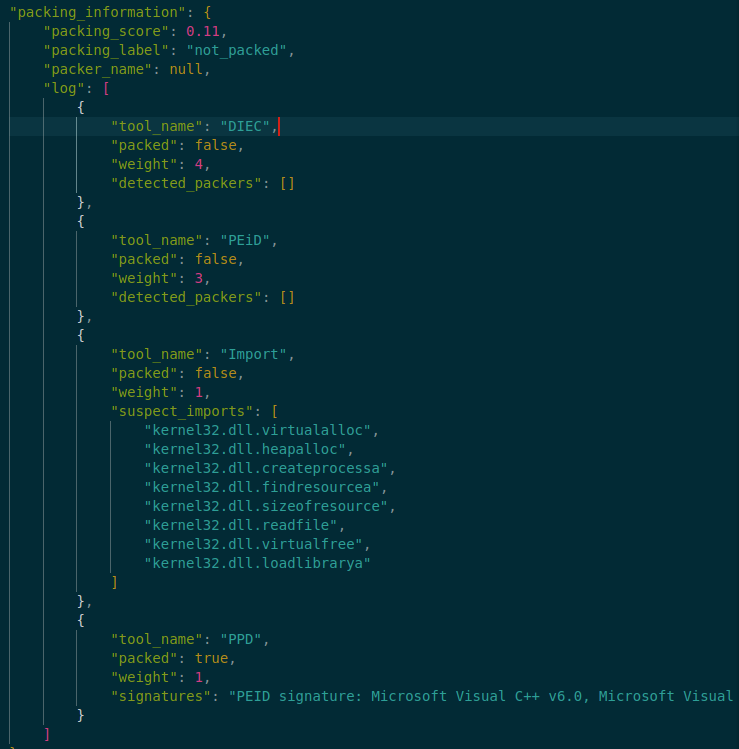}
    \caption{Packing signatures}
    \label{fig:section_packing}
\end{figure}

The fourth section, illustrated in Figure \ref{fig:section_packing} is related to the \textbf{obfuscation signatures}, in particular if the file is packed. Packing code is a common technique used by threat actor and malware editors to hide malicious payload and disable or prevent static analysis \cite{VignaGiovanni}. Even if it is a technique used by software editors for intellectual property reasons, it is also a indicator of suspicious for malicious file analysis. 

We use open-source tools as DIEC\footnote{https://github.com/horsicq/Detect-It-Easy}, PEid\footnote{https://github.com/packing-box/peid} or PyPackerDetecte\footnote{https://github.com/packing-box/pypackerdetect} to detect signatures of packer tools. Based on these different tools, we extract signatures of potential packers, and other indications of potential packing. We define the packing label $l_p$ as following: 
\begin{equation}
    l_p = \frac{\sum_{i=1}^n w_il_{pi}}{\sum_{i=1}^nw_i}
\end{equation}
where $l_{pi} \in \{0, 1\}$ is the packing label returned by each packing detection tool, with $l_{pi}=1$ if the tool detects a packer signature or indication of packing. The weights $w_i$ are defined by the efficiency of the packing detectors. Based on the final signature , we also return the potential packing software used on the file.

Then the last section is related to \textbf{Mitre Att\&ck and MBC identification}. The MITRE ATT\&CK\footnote{https://attack.mitre.org/} framework is a globally accessible knowledge base of adversary tactics and techniques based on real-world observations. It provides a structured approach to understanding how cyberthreats operate across the attack lifecycle. For malware analysis, especially in categorizing threats like trojans and ransomware, ATT\&CK helps analysts to map malicious behaviors to specific techniques such as credential dumping, command and control, or data encryption for impact. By aligning observed malware behaviors with ATT\&CK techniques, security professionals can gain deeper insight into the malware's functionality, objectives, and potential mitigation. This behavioral mapping is critical for threat intelligence, detection engineering, and incident response, allowing defenders to anticipate attacker moves and improve their defensive posture.

The Malware Behavior Catalog (MBC)\footnote{https://github.com/MBCProject/mbc-markdown} is a structured framework developed to describe and classify the behaviors of malware in a consistent and standardized way. Unlike general threat frameworks, MBC focuses specifically on how a malware operates, including capabilities such as data theft, persistence, evasion, and execution. It complements MITRE ATT\&CK by diving deeper into malware-centric techniques and patterns, making it especially valuable for categorizing malware categories like trojans, ransomware, and spyware. By mapping observed behaviors to MBC entries, analysts can better understand a malware sample’s functionality, compare it with known behaviors, and enrich threat intelligence. This detailed behavioral insight supports reverse engineering, detection development, and attribution efforts, enhancing the overall malware analysis process.

\begin{figure}[!h]
    \centering
    \includegraphics[width=0.7\linewidth]{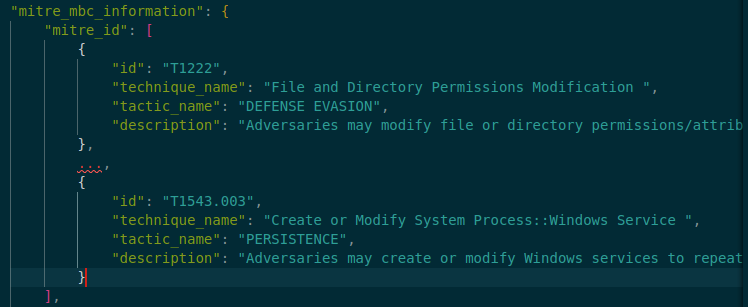}
    \caption{MITRE ATT\&CK and MBC}
    \label{fig:section_mbc}
\end{figure}

Using CAPA\footnote{https://github.com/mandiant/capa}, we extract MITRE ATT\&CK and MBC information and integrate it into our JSON report, as described in Figure \ref{fig:section_mbc}. The purpose of including this information is to provide to malware analyst behavioral information about a malicious file, and to improve malware analysis by deep learning models by giving a better description of the files.


\section{Experiments}
\label{sec:Experiments}

In this section, we present the results of our malware classification approach based on transformers architecture. This is a much more complicated use-case than malware detection. We do not address the detection task here as we believe it does not require the use of transformers, and it has been addressed previously in other works with good results \cite{anderson2018EMBER,Gaber/Ahmed,marais2022aibasedmalwareransomwaredetection}.

\subsection{Models and training }

We have chosen the BERT base uncased model \cite{devlin2019bertpretrainingdeepbidirectional}, a transformers model pretrained on a large corpus of English data in a self supervised fashion using a masked modeling objective, because it is a very versatile model with only 110 million parameters. One limitation of Bert is in the number of $512$ tokens, but by selecting the most relevant parts of the reports, we end up with an encoding where the average number of tokens is $403$, the median number is $350$ and only $15 \%$  of the data is greater than $512$ tokens. We consider this a good compromise, given the quality of the Bert model and its speed of training.
To understand what this restriction implies, we also test the ModernBert model, a similar model with a limit of 1024 tokens. Only $1.45 \%$ of the data is greater than $1024$ token.
Using these small models enables us to perform our tests on an Apple M4 Pro macbook. We also did some tests by fine-tuning deepseek models with 1.5B parameters on a $4090$ RTX, but the computation time was too long and the results not so satisfying.

With BERT models, we are able to process and analyze JSON reports associated to each malware. Transformers, with their ability to capture complex contextual relationships in textual data, are a very well-adapted solution for this classification problem.
The following results highlight the effectiveness of this approach and provide insights into the strengths of transformer-based malware classification, if we have the adapted preprocessing. As the classes are not balanced, we focus on the F1-score, but we give other metrics for information. The train/test split is as usual 80/20, and we train our model on 10 epochs with a batch size of 16. We classify 8 categories, ranked from 0 to 7, in the following order: trojan, worm, ransomware, backdoor, information stealer, downloader, dropper, and virus. 

\subsection{BERT Results}

\begin{figure}[!h]
    \centering
    \includegraphics[width=0.8\linewidth]{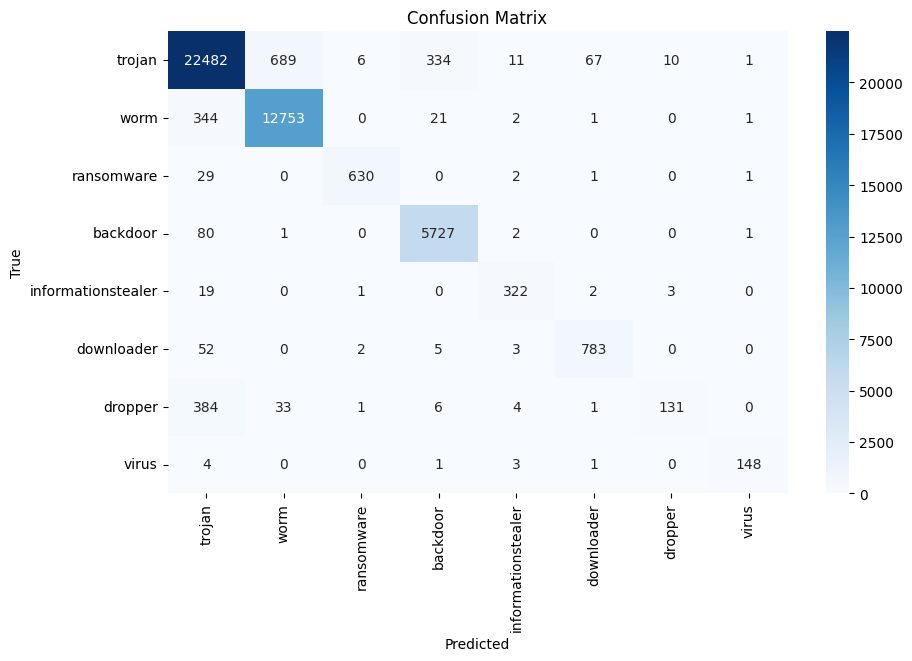}
    \caption{Confusion matrix for the train dataset.}
    \label{fig:matrix-train}
\end{figure}

\begin{figure}[!h]
    \centering
    \includegraphics[width=0.8\linewidth]{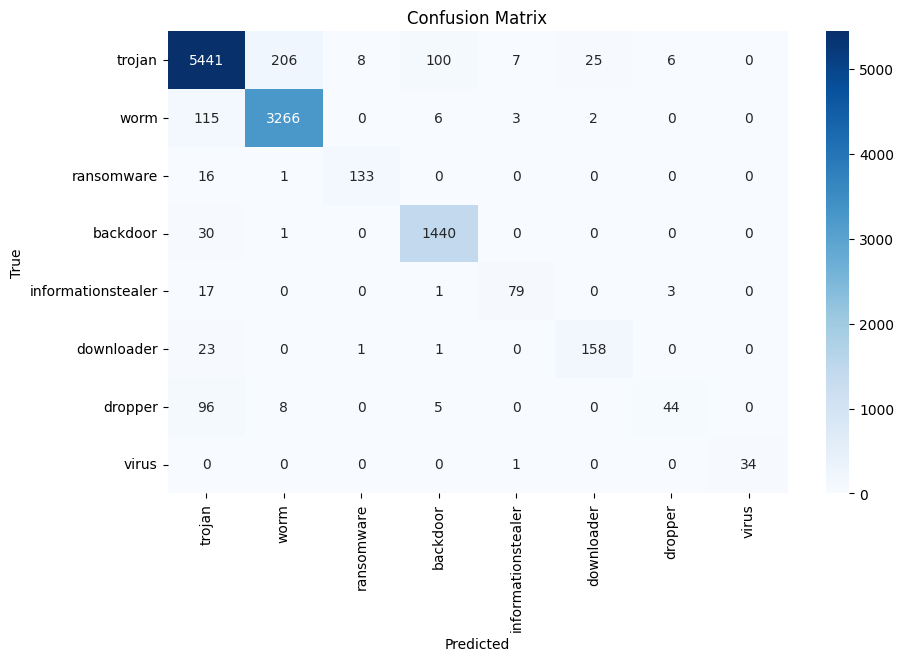}
    \caption{Confusion matrix for the test dataset. }
    \label{fig:matrix-test}
\end{figure}

In order to give a first illustration of our results concerning the classification task with BERT uncase model, we compute the confusion matrices for the train set in Figure \ref{fig:matrix-train} and for the test set in Figure \ref{fig:matrix-test}. We can see that the model is really good at identifying the most represented categories. It is less clear for the under-represented categories, but the results are still good. The only category where the classification seems not to be efficient is the dropper category, where the model tends to identify droppers as trojans. It is not very surprising, and can be explained because droppers (also named trojan-dropper) are a type of trojan that has been designed to install a malware into a computer. For example, F-Secure or Kaspersky categorize it as trojan-dropper.

To get a more precise idea of our model's performances, we present in Tables \ref{tab:report-train} and \ref{tab:report-test} the accuracy, F1-score and recall of the train and test sets respectively. We also specify the number of sample of each class in these tables, and compute the average scores of the classification task. 

\begin{table}[!h]
\centering
\caption{Training classification report with BERT}
\label{tab:report-train}
\begin{tabular}{|c|c|c|c|c|}
\hline
\textbf{Class} & \textbf{Accuracy} & \textbf{Recall} & \textbf{F1-score} & \textbf{Support} \\
\hline
trojan & 0.96 & 0.95 & 0.96 & 23600 \\
worm & 0.95 & 0.97 & 0.96 & 13122 \\
ransomware & 0.98 & 0.95 & 0.97 & 663 \\
backdoor & 0.94 & 0.99 & 0.96 & 5811 \\
infostealer & 0.92 & 0.93 & 0.93 & 347 \\
downloader & 0.91 & 0.93 & 0.92 & 845 \\
dropper & 0.91 & 0.23 & 0.37 & 560 \\
virus & 0.97 & 0.94 & 0.96 & 157 \\
\hline
\textbf{Global} & & & & \\
\hline
\textbf{Accuracy} & & & 0.95 & 45105 \\
\textbf{Macro avg} & 0.94 & 0.86 & 0.88 & 45105 \\
\textbf{Weighted avg} & 0.95 & 0.95 & 0.95 & 45105 \\
\hline
\end{tabular}
\end{table}

\begin{table}[!h]
\centering
\caption{Testing classification report with BERT}
\label{tab:report-test}
\begin{tabular}{|c|c|c|c|c|}
\hline
\textbf{Class} & \textbf{Accuracy} & \textbf{Recall} & \textbf{F1-score} & \textbf{Support} \\
\hline
trojan & 0.95 & 0.94 & 0.94 & 5793 \\
worm & 0.94 & 0.96 & 0.95 & 3392 \\
ransomware & 0.94 & 0.89 & 0.91 & 150 \\
backdoor & 0.93 & 0.98 & 0.95 & 1471 \\
infostealer & 0.88 & 0.79 & 0.83 & 100 \\
downloader & 0.85 & 0.86 & 0.86 & 183 \\
dropper & 0.83 & 0.29 & 0.43 & 153 \\
virus & 1.00 & 0.97 & 0.99 & 35 \\
\hline
\textbf{Global} & & & & \\
\hline
\textbf{Accuracy} & & & 0.94 & 11277 \\
\textbf{Macro avg} & 0.91 & 0.84 & 0.86 & 11277 \\
\textbf{Weighted avg} & 0.94 & 0.94 & 0.94 & 11277 \\
\hline
\end{tabular}
\end{table}

The classification results obtained using the BERT model demonstrate strong performances across most malware categories, even the under-represented, with a F1-score greater than $0.92$ (resp. $0.83$) for all classes except one during the training (resp. testing). As illustrated by the Tables \ref{tab:report-train} and \ref{tab:report-test}, the model achieves very high accuracy and F1-score in distinguishing all classes except droppers with minimal misclassification. Unsurprisingly, under-represented categories are harder to identify for the model, but it could be fixed if we manage to increase the number of samples from these categories.  

\subsection{ModernBERT results}
\begin{figure}[!h]
    \centering
    \includegraphics[width=0.8\linewidth]{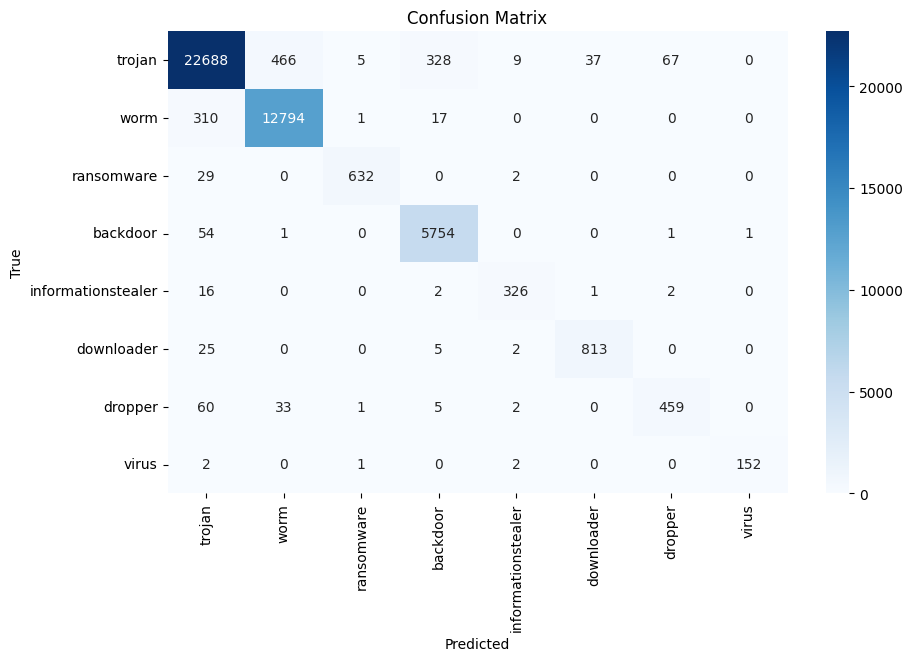}
    \caption{Confusion matrix for the train dataset with ModernBERT.}
    \label{fig:Modern-matrix-train}
\end{figure}

\begin{figure}[!h]
    \centering
    \includegraphics[width=0.8\linewidth]{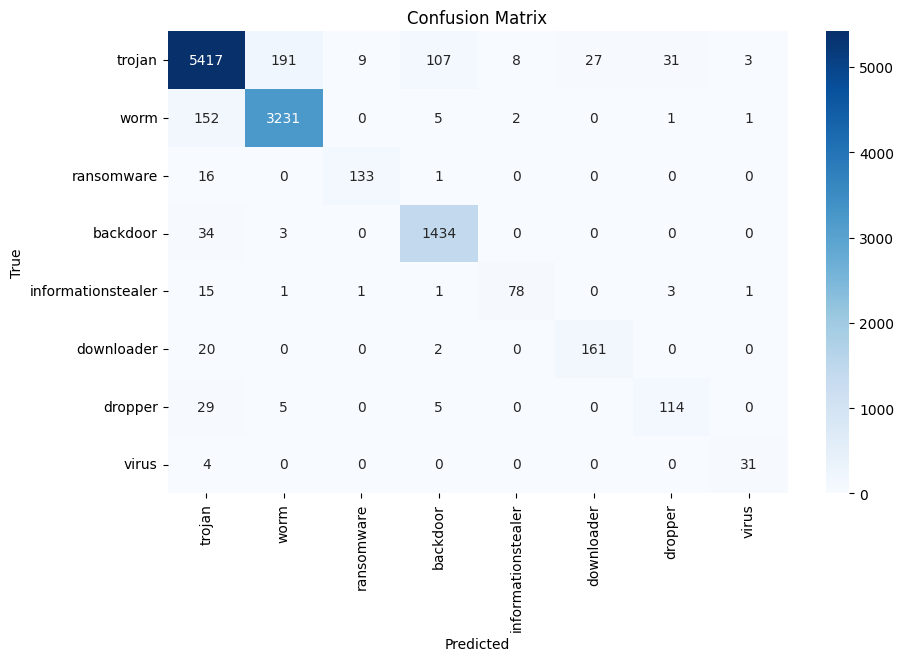}
    \caption{Confusion matrix for the test dataset with ModernBERT. }
    \label{fig:Modern-matrix-test}
\end{figure}

In this section, we first illustrate the results of our classification task using ModernBERT. In this aim Figure \ref{fig:Modern-matrix-train} shows the confusion matrix for the train dataset, while Figure \ref{fig:Modern-matrix-test} shows the same confusion matrix, but for the test dataset. 

Increasing the maximum input length for the encoder with the ModernBERT improves classification performance for certain malware families. While the classification of well-represented families remains strong, extending the sequence length significantly enhances the detection of droppers, which was a limitation of the original BERT model. Specifically, the F1-score for droppers increases from 0.37 to 0.84 on the training set according to Table \ref{tab:report-train-Modern}, and from 0.43 to 0.75 on the test set according to Table \ref{tab:report-test-Modern}. Although the tables may suggest a decrease in performance for viruses (from 0.99 to 0.87 between Tables \ref{tab:report-test} and \ref{tab:report-test-Modern}), this actually corresponds to only three additional errors due to the small number of virus samples.

\begin{table}[!h]
\centering
\caption{Training classification report with ModernBERT}
\label{tab:report-train-Modern}
\begin{tabular}{|c|c|c|c|c|}
\hline
\textbf{Classe} & \textbf{Précision} & \textbf{Rappel} & \textbf{F1-score} & \textbf{Support} \\
\hline
trojan & 0.98 & 0.96 & 0.97 & 23600 \\
worm & 0.96 & 0.98 & 0.97 & 13122 \\
ransomware & 0.99 & 0.95 & 0.97 & 663 \\
backdoor & 0.94 & 0.99 & 0.97 & 5811 \\
infostealer & 0.95 & 0.94 & 0.94 & 347 \\
downloader & 0.96 & 0.96 & 0.96 & 845 \\
dropper & 0.87 & 0.82 & 0.84 & 560 \\
virus & 0.99 & 0.97 & 0.98 & 157 \\
\hline
\textbf{Global} & & & & \\
\hline
\textbf{Accuracy} & & & 0.97 & 45105 \\
\textbf{Macro avg} & 0.95 & 0.95 & 0.95 & 45105 \\
\textbf{Weighted avg} & 0.97 & 0.97 & 0.97 & 45105 \\
\hline
\end{tabular}
\end{table}

\begin{table}[!h]
\centering
\caption{Testing classification report with ModernBERT}
\label{tab:report-test-Modern}
\begin{tabular}{|c|c|c|c|c|}
\hline
\textbf{Classe} & \textbf{Précision} & \textbf{Rappel} & \textbf{F1-score} & \textbf{Support} \\
\hline
trojan & 0.95 & 0.94 & 0.94 & 5793 \\
worm & 0.94 & 0.95 & 0.95 & 3392 \\
ransomware & 0.93 & 0.89 & 0.91 & 150 \\
backdoor & 0.92 & 0.97 & 0.95 & 1471 \\
infostealer & 0.89 & 0.78 & 0.83 & 100 \\
downloader & 0.86 & 0.88 & 0.87 & 183 \\
dropper & 0.77 & 0.75 & 0.75 & 153 \\
virus & 0.86 & 0.89 & 0.87 & 35 \\
\hline
\textbf{Global} & & & & \\
\hline
\textbf{Accuracy} & & & 0.94 & 11277 \\
\textbf{Macro avg} & 0.89 & 0.88 & 0.88 & 11277 \\
\textbf{Weighted avg} & 0.94 & 0.94 & 0.94 & 11277 \\
\hline
\end{tabular}
\end{table}

Using the ModernBERT model to have a maximum sequence length from $512$ to $1024$ does not noticeably affect training speed, but it does require more memory. Nevertheless, this trade-off appears justified given the improved malware classification results. Overall, these results highlight the potential of leveraging language models associated with very complete reports on malware for automated classification. It is thus important to create such reports, gathering as much information as possible, from static and behavioral analysis.  

\subsection{Multi-CNN classification}

To benchmark our results, we use a multi-convolutional neural network defined in
\cite{quertier2024usemulticnnssectionanalysis}, where we used Grayscale's method \cite{nataraj2011malware} to transform the different sections of the PE files in images, and then train one CNN per section before gathering all the results thanks to a scoring function. Each CNN contains five layers, and is optimized with Adam \cite{kingma2017adammethodstochasticoptimization}, with a batch size of 32.  This model is fast to train and the preprocessing is now a very common while intuitive tool to use, so it is interesting to see if this ``naive'' model has already good performances or if we indeed need to work with a more elaborated model. 

\begin{figure}[!h]
    \centering
    \includegraphics[width=0.8\linewidth]{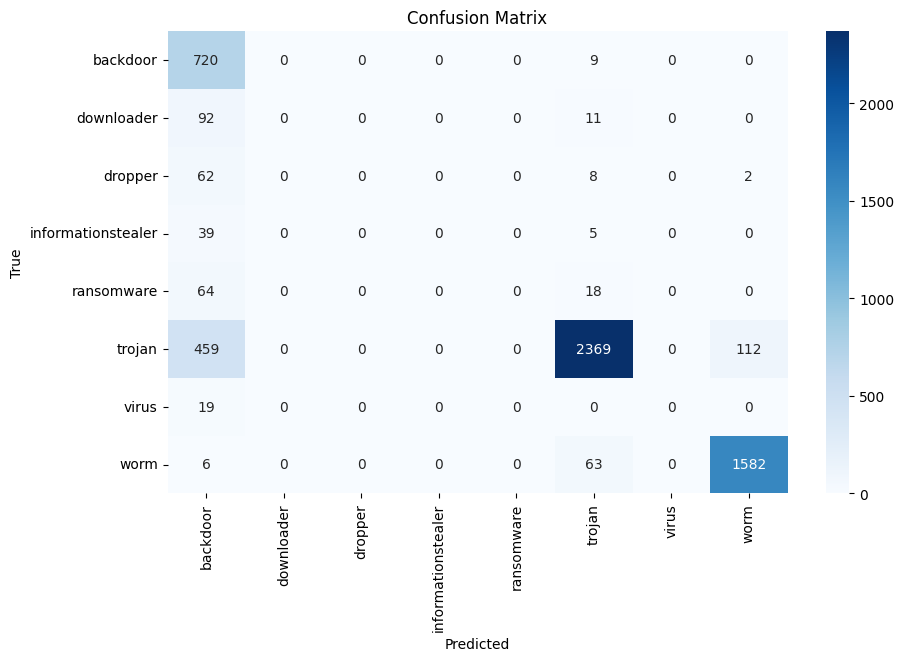}
    \caption{Confusion matrix for the test dataset wit the multi-CNN.}
    \label{fig:matrix-cnn-test}
\end{figure}

Here we present the confusion matrix in Figure \ref{fig:matrix-cnn-test} and the classification metrics in Table \ref{tab:cnn-report-test} only for the test dataset. 

\begin{table}[!h]
\centering
\caption{Testing classification report with multi-CNN}
\label{tab:cnn-report-test}
\begin{tabular}{|c|c|c|c|c|}
\hline
\textbf{Class} & \textbf{Accuracy} & \textbf{Recall} & \textbf{F1-score} & \textbf{Support} \\
\hline
backdoor & 0.49 & 0.99 & 0.66 & 729 \\
downloader & 0.00 & 0.00 & 0.00 & 103 \\
dropper & 0.00 & 0.00 & 0.00 & 72 \\
infostealer & 0.00 & 0.00 & 0.00 & 44 \\
ransomware & 0.00 & 0.00 & 0.00 & 82 \\
trojan & 0.95 & 0.81 & 0.87 & 2940 \\
virus & 0.00 & 0.00 & 0.00 & 19 \\
worm & 0.93 & 0.96 & 0.95 & 1651 \\
\hline
\textbf{Global} & & & & \\
\hline
\textbf{Accuracy} & & & 0.83 & 5640 \\
\textbf{Macro avg} & 0.30 & 0.34 & 0.31 & 5640 \\
\textbf{Weighted avg} & 0.83 & 0.83 & 0.82 & 5640 \\
\hline
\end{tabular}
\end{table}

We can see  in Figure \ref{fig:matrix-cnn-test} that all the under-represented categories are labeled as  backdoors, showing the incapacity of the model to learn correctly on an unbalanced dataset. In Table \ref{tab:cnn-report-test}, the accuracy is decent only for trojan and worms, which are the two most represented classes of the dataset. 

\subsection{Balanced Dataset}

\begin{figure}[!h]
    \centering
    \includegraphics[width=0.8\linewidth]{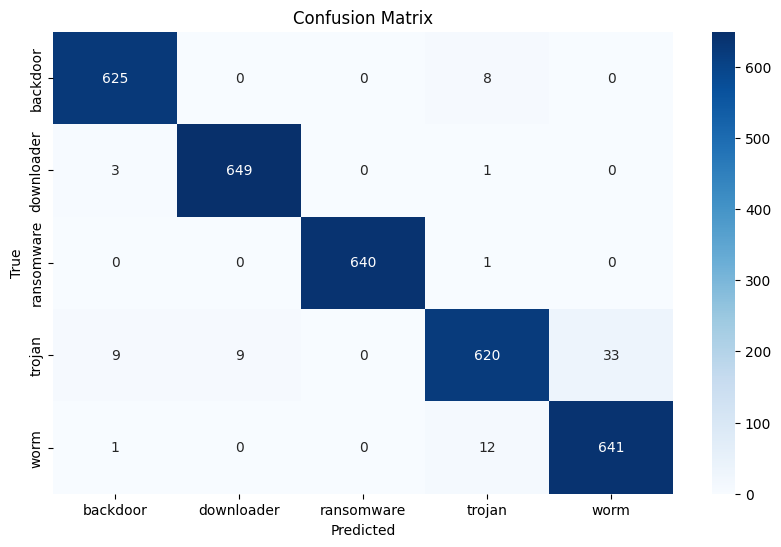}
    \caption{Confusion matrix for the balanced train dataset with ModernBERT.}
    \label{fig:Modern-balanced-train}
\end{figure}

\begin{figure}[!h]
    \centering
    \includegraphics[width=0.8\linewidth]{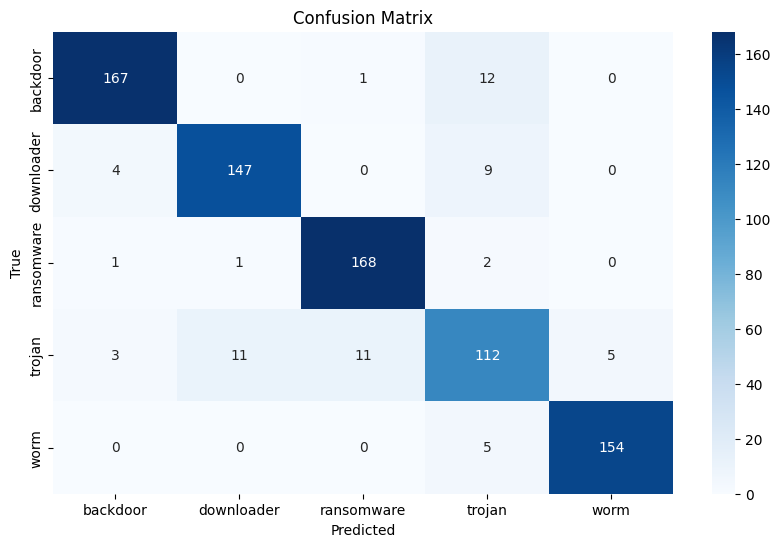}
    \caption{Confusion matrix for the balanced test dataset with ModernBERT.}
    \label{fig:Modern-balanced-test}
\end{figure}

In order to complete our tests, we run our classification task on a balanced dataset. We isolate the five more represented classes - namely trojans, worms, backdoors, downloaders and ransomwares - and take 813 samples from each of them. As usual, we use a train/test split of 80/20. We only test ModernBERT on this dataset, because here we just want to show that the previous unbalanced dataset does not bias the accuracy results. Results are illustrated in Table \ref{tab:class_balanced_train} for the train set and in Table \ref{tab:class_balanced_test} for the test set. We also provide the confusion matrices in Figures \ref{fig:Modern-balanced-train} and \ref{fig:Modern-balanced-test}. 

As we can see in Table \ref{tab:class_balanced_test}, the smallest accuracy is 0.80, and concerns the downloader class. These results show that even with a balanced dataset our semantic preprocessing is well-adapted to a classification task using LLMs. 

\begin{table}[ht]
\centering
\caption{Training classification report on balanced dataset with ModernBERT}
\label{tab:class_balanced_train}
\begin{tabular}{|c|c|c|c|c|}
\hline
\textbf{Classe} & \textbf{Accuracy} & \textbf{Recall} & \textbf{F1-score} & \textbf{Support} \\
\hline
backdoor & 0.98 & 0.99 & 0.98 & 633 \\
downloader & 0.99 & 0.99 & 0.99 & 653 \\
ransomware & 1.00 & 1.00 & 1.00 & 641 \\
trojan & 0.97 & 0.92 & 0.94 & 671 \\
worm & 0.95 & 0.98 & 0.97 & 654 \\
\hline
\textbf{Global} & & & & \\
\hline
\textbf{Accuracy} & & & 0.98 & 3252 \\
\textbf{Macro avg} & 0.98 & 0.98 & 0.98 & 3252 \\
\textbf{Weighted avg} & 0.98 & 0.98 & 0.98 & 3252 \\
\hline
\end{tabular}
\end{table}

\begin{table}[ht]
\centering
\caption{Testing classification report on balanced dataset with ModernBERT}
\label{tab:class_balanced_test}
\begin{tabular}{|c|c|c|c|c|}
\hline
\textbf{Classe} & \textbf{Accuracy} & \textbf{Recall} & \textbf{F1-score} & \textbf{Support} \\
\hline
backdoor & 0.95 & 0.93 & 0.94 & 180 \\
downloader & 0.92 & 0.92 & 0.92 & 160 \\
ransomware & 0.93 & 0.98 & 0.95 & 172 \\
trojan & 0.80 & 0.79 & 0.79 & 142 \\
worm & 0.97 & 0.97 & 0.97 & 159 \\
\hline
\textbf{Global} & & & & \\
\hline
\textbf{Accuracy} & & & 0.92 & 813 \\
\textbf{Macro avg} & 0.92 & 0.92 & 0.92 & 813 \\
\textbf{Weighted avg} & 0.92 & 0.92 & 0.92 & 813 \\
\hline
\end{tabular}
\end{table}

\newpage ~ \newpage

\section*{Conclusion and future works}

In a nutshell, we present in this article a new preprocessing for PE malware files, which allows semantic analysis via Large Languages Models. Our preprocessing generates JSON reports gathering static and behavioral features, such as packer signatures and YARA rules. Our preprocessing gathers concrete features, that can be read and understood directly by analysts.  We used a BERT model in order to classify eight categories of malware, on a realistic dataset where some categories are under-represented. The classification results are very good and encouraging, showing that our preprocessing is suitable for semantic analysis.

This work opens several perspectives to enhance our performances. First, static and behavioral analysis would benefit to be completed by dynamic analysis, which will be incorporated in the reports, giving an even more complete preprocessing for PE files. We also want to refine our classification by detecting the category but also the family of a malware. In this aim it will be suitable to increase the size of our dataset, gathering more samples from the diverse categories and thus the families. This will probably increase the performances of our model, which might be able to differentiate more precisely close categories such as droppers and trojans for example. We also might be able to classify more categories and more families. Finally, the big advantage of LLMs is the use of attention mechanisms. We ran some tests in this direction but it was not conclusive for our classification use-case, but it will be interesting to use it for instance to summarize the behavior of a sample based on its report. For the classification use-case, we will try to provide some explainability based of our knowledge concerning the information contained in the report: for example some Mitre Att\&ck id may be specific to the ransomware category. 

Note that our feature extraction and report generation method is currently a prototype which will be enhanced according to the previous perspectives, and released publicly for open research. We believe that this new complete preprocessing is a very good tool for AI-assisted Malware classification, as it gathers in the same place all the information about a file, allowing very modular analysis if someone wants to use only specific features. We expect this contribution to help solving the issue of bad data representation for Malware classification.


\bibliographystyle{splncs04}
\bibliography{bib}

\end{document}